\documentclass[preprint,pra,amsmath,amssymb]{revtex4}
\usepackage{graphicx}% Include figure files
\usepackage{dcolumn}% Align table columns on decimal point
\usepackage{epic}%
\usepackage{eepic}%
\usepackage{bm}% bold math
\begin{document}

%\preprint{APS/123-QED}

%\newcommand{\zlabel}[1]{\label{#1} \mbox{\tiny (#1)}} 
\newcommand{\zlabel}{\label} 
\newcommand{\hH}{\hat{H}} %\mbox{\tiny hH:$\hH$}
\newcommand{\hT}{\hat{T}} %\mbox{\tiny hT:$\hT$}
\newcommand{\ke}[1]{-\mbox{$\frac{1}{2}$}\nabla_{{#1}}^2} %\mbox{\tiny ke\{x\}:$\ke{x}$} 
\newcommand{\roo}{{\rho_1}} %\mbox{\tiny roo:$\roo$}
\newcommand{\hroo}{{\hat{\rho}_1}} %\mbox{\tiny hroo:$\hroo$}
\newcommand{\hvroo}{{\hat{\varrho}_1}} %\mbox{\tiny hvroo:$\hvroo$}
\newcommand{\vroo}{{\varrho}_1} %\mbox{\tiny vroo:$\vroo$}
\newcommand{\kp}{\kappa} %\mbox{\tiny kp:$\kp$}
\newcommand{\hkp}{\hat{\kappa}} %\mbox{\tiny hkp:$\hkp$}
\newcommand{\rc}[1]{r^{-1}_{#1}} %\mbox{\tiny rc:$\rc{x}$}
\newcommand{\fc}{\frac} 
\newcommand{\lt}{\left} 
\newcommand{\rt}{\right} 
\newcommand{\ran}{\rangle} 
\newcommand{\lan}{\langle} 
\newcommand{\cE}{{\cal E}} 
\newcommand{\hVee}{\hat{V}_{\text{ee}}} 
\newcommand{\hV}{\hat{V}} 
\newcommand{\hGm}{\hat{\Gamma}}
\newcommand{\Gm}{\Gamma}
\newcommand{\gm}{\gamma}
\newcommand{\si}{\sigma}
\newcommand{\dg}{\dagger} 
\newcommand{\mr}{\mathbf{r}}
\newcommand{\mx}{\mathbf{x}}
\newcommand{\om}{\omega}
\newcommand{\hE}{\hat{E}}
\newcommand{\dt}{\delta}
\newcommand{\al}{\alpha} %\mbox{\tiny al:$\al$}
\newcommand{\be}{\beta} %\mbox{\tiny be:$\be$}
\newcommand{\hs}[1]{\hspace{#1ex}}
\newcommand{\vs}[1]{\vspace{#1ex}}
\newcommand{\pr}{\prime}
\newcommand{\hvj}{\hat{v}_{J}} %\mbox{\tiny hvj:$\hvj$}
\newcommand{\vj}{v_{J}} %\mbox{\tiny vj:$\vj$}
\newcommand{\hvxc}{\hat{v}_{\text{xc}}} %\mbox{\tiny hvxc:$\hvxc$}
\newcommand{\vxc}{v_{\text{xc}}} %\mbox{\tiny hvxc:$\vxc$}
\newcommand{\hvx}{\hat{v}_{\text{x}}} %\mbox{\tiny hvx:$\hvx$}
\newcommand{\hvco}{\hat{v}_{\text{co}}} %\mbox{\tiny hvco:$\hvco$}
\newcommand{\vco}{v_{\text{co}}} %\mbox{\tiny vco:$\vco$}
\newcommand{\hvcoroo}{\hat{v}_{\text{co}}^{\rho_1}} %\mbox{\tiny hvcoroo:$\hvcoroo$}
\newcommand{\vcoroo}{v_{\text{co}}^{\rho_1}} %\mbox{\tiny vcoroo:$\vcoroo$}
\newcommand{\hvxroo}{\hat{v}_{\text{x}}^{\rho_1}} %\mbox{\tiny hvxroo:$\hvxroo$}
\newcommand{\vxroo}{v_{\text{x}}^{\rho_1}} %\mbox{\tiny vxroo:$\vxroo$}
\newcommand{\hvecroo}{\hat{v}_{\text{ec}}^{\rho_1}} %\mbox{\tiny hvecroo:$\hvecroo$}
\newcommand{\hvec}{\hat{v}_{\text{ec}}} %\mbox{\tiny hvec:$\hvec$}
\newcommand{\vecroo}{v_{\text{ec}}^{\rho_1}} %\mbox{\tiny vecroo:$\vecroo$}
\newcommand{\tPsi}{\tilde{\Psi}}
\newcommand{\eps}{\epsilon} %\mbox{\tiny eps:$\eps$}
\newcommand{\hF}{\hat{F}} %\mbox{\tiny hF:$\hF$}
\newcommand{\cF}{{\cal F}} %\mbox{\tiny cF:$\cF$}
\newcommand{\hcF}{{\cal \hat{F}}} %\mbox{\tiny hcF:$\hcF$}
\newcommand{\hv}{\hat{v}} %\mbox{\tiny hv:$\hv$}
\newcommand{\hw}{\hat{w}} %\mbox{\tiny hw:$\hw$}
\newcommand{\nn}{\nonumber} 
\newcommand{\rco}{r_{12}^{-1}} %\mbox{\tiny rco:$\rco$}
\newcommand{\Eco}{E_{\text{co}}} %\mbox{\tiny Eco:$Eco$}
\newcommand{\Ex}{E_{\text{x}}} %\mbox{\tiny Ex:$Ex$}
\newcommand{\tEco}{\tilde{E}_{\text{co}}} %\mbox{\tiny tEco:$tEco$}
\newcommand{\cEco}{{\cal E}_{\text{co}}} %\mbox{\tiny Eco:$Eco$}
\newcommand{\vep}{\varepsilon} %\mbox{\tiny vep:$\vep$}

\title{Expressions for the Exchange Correlation Potential and Exchange--Correlation
Functional of Kohn--Sham Density Functional Theory}

\author{James P. Finley}
\email{james.finley@enmu.edu}

\affiliation{
Department of Physical Sciences,
Eastern New~Mexico University,
Station \#33, Portales, NM 88130}
\affiliation{
Department of Applied Chemistry,
Graduate School of Engineering,
The University of Tokyo,
7-3-1 Hongo, Bunkyo-ku,
Tokyo, 113-8656 Japan}
\date{\today}

\begin{abstract}

The State--Specific Kohn--Sham Density Functional Theory [arXiv:physics/0506037]
is used to derive the Kohn-Sham exchange-correlation potential $\vxc$ and
exchange-correlation functional $E_{\text{xc}}$ as explicit functionals of $v_s$ and
$\varphi_1$, where $v_s$ is the local, one-body potential from the Kohn--Sham
equations, and $\varphi_1$ is the spinless one-particle density matrix from the
Kohn--Sham noninteracting state, say $|\varphi_1\ran$. In other words,
$|\varphi_1\ran$ is the ground state eigenfunction of the noninteracting
Schr\"odinger equation with the one-body potential $v_s$.  For simplicity, we only
consider noninteracting states that are closed-shell states and interacting states
that are nondegenerate, singlet ground-states.

\end{abstract}

\maketitle

%06/10/05 Tokyo Time

\section{Introduction}

The Kohn-Sham version of density functional theory plays a major role in both
quantum chemistry and condensed matter physics
\cite{Dreizler:90,Parr:89,Springborg:97,Ellis:95,Gross:94,Seminario:95,Handy:97}.
Unfortunately, the exchange-correlation functional $E_{\text{xc}}$ is an unknown,
implicit functional, and there is no systematic method to improve approximations.
Recently we have derived a generalization of the Kohn--Sham approach in which the
correlation energy $\Eco$ is assumed to be an explicit functional of $v$ and
$\roo$, where $v$ is the external potential from the interacting target-state, and
$\roo$ is the spinless one-particle density matrix from the noninteracting states
\cite{Finley:ssks.arxiv,Finley:ssks}.  In this approach, errors from Coulomb
self-interactions do not occur, nor the need to introduce functionals defined by a
constraint search.  Furthermore, the exchange energy $\Ex$ is treated as in
Hartree--Fock theory, as an explicit functional of $\roo$. Below, we use this
approach to derive the Kohn-Sham exchange-correlation potential $\vxc$ and
exchange-correlation functional $E_{\text{xc}}$ as explicit functionals of $v_s$
and $\varphi_1$, where $v_s$ is the local, one-body potential from the Kohn--Sham
equations, and $\varphi_1$ is the one-particle density matrix from the Kohn--Sham
noninteracting state, say $|\varphi_1\ran$. In other words, $|\varphi_1\ran$ is
the ground state eigenfunction of the noninteracting Schr\"odinger equation with
the one-body potential $v_s$.

\section{State-specific Kohn--Sham density functional theory}

In state-specific Kohn--Sham density functional theory
\cite{Finley:ssks.arxiv,Finley:ssks}, we use an energy functional $E_v[\roo]$ that is
assumed to be an explicit functional of the external potential $v$ and the spinless
one-particle density matrix $\roo$, where $\roo$ comes from a
closed-shell determinantal state, say $|\roo\ran$; this energy functional is given
by
\begin{eqnarray} \zlabel{3819}
E_v[\roo] = 
\fc{\lan \tPsi_{v\roo}|\hH_{v}|\tPsi_{v\roo}\ran}
   {\lan \tPsi_{v\roo}|\tPsi_{v\roo}\ran},
\end{eqnarray}
where the trial wave function $\tPsi_{v\roo}$ generates the exact, or target, wave
function, say $\Psi_{n}$, under the following conditions:
\begin{eqnarray} \zlabel{3810}
|\tPsi_{v\vroo}\ran = |\Psi_{n}\ran; 
\;\; \vroo \longrightarrow  n,\;\; n \longrightarrow N,v,
\end{eqnarray}
where $n$ is the electron density of $\Psi_{n}$, and $\Psi_{n}$ is the ground state
singlet eigenfunction of $\hH_{v}$.  The right side notation of Eq.~(\ref{3810})
indicates that $\vroo$ is a spin-less one particle density matrix that delivers
the density $n$; according to the Hohenberg-Kohn theorem
\cite{Hohenberg:64a,Parr:89,Dreizler:90}, $n$ also determines the external
potential $v$, and $n$ determines the number of electrons $N$. Using
Eqs.~(\ref{3819}) and (\ref{3810}), we have
\begin{eqnarray} \zlabel{8273}
E_v[\vroo] = \cE_{n}, \;\; \vroo \longrightarrow n, 
\;\; n \longrightarrow N,v.
\end{eqnarray}
Here, $\cE_{n}$ is the exact electronic energy of the target state:
\begin{eqnarray} \zlabel{0283}
\hH_{v} |\Psi_{n}\ran = \cE_{n}|\Psi_{n}\ran; \;\; n \longrightarrow N,v.
\end{eqnarray}
where the Hamiltonian operator is given by
\begin{eqnarray} \zlabel{8793}
\hH_{v} = \hT + \hVee + \hV_v,
\end{eqnarray}
and we have
\begin{eqnarray}
\hT   &=& \sum_{i}^{N}(\ke{i}), \\
\hVee &=& \fc12 \sum_{i\ne j}^{N}\rc{ij}, \\ 
\hV_v &=& \sum_{i}^{N}v(i).
\end{eqnarray}

The electronic energy functional can also be written as
\begin{eqnarray} \label{1272}
E_v[\roo] = \int d\mr_1\, \lt[\ke{1}\roo(\mr_1,\mr_2)\rt]_{\mr_2=\mr_1}
+ \int d\mr\, v(\mr) \rho_s(\mr) 
\hs{30}
\nn \\ \mbox{}
\hs{10}
+ E_J[\rho_s] + \Ex[\roo] + \Eco[\roo,v]
+ \int d\mr\, v(\mr) \tilde{\rho}_c(\mr),
\end{eqnarray}
where the Coulomb and exchange energies are given by the following:
\begin{eqnarray} %\zlabel{}
E_J[\rho_s] &=& \fc12 \int \int \rco d\mr_1 d\mr_2 \rho(\mr_1)\, \rho(\mr_2), \\ 
-\Ex[\roo] &=& \fc14 \int \int \rco d\mr_1 d\mr_2 \roo(\mr_1,\mr_2)\,
\roo(\mr_2,\mr_1),
\end{eqnarray}
and the correlation-energy functional is
\begin{eqnarray} \zlabel{1928}
\Eco[\roo,v] =  
\fc{\lan \tPsi_{v\roo}|\hT|\tPsi_{v\roo}\ran}
   {\lan \tPsi_{v\roo}|\tPsi_{v\roo}\ran}
-  \lan \roo|\hT|\roo \ran
+ \fc{\lan \tPsi_{v\roo}|\hVee|\tPsi_{v\roo}\ran}
   {\lan \tPsi_{v\roo}|\tPsi_{v\roo}\ran}
- \lan \roo|\hVee|\roo \ran.
\end{eqnarray}
Furthermore, $\tilde{\rho}_c$ is the correlation density of the trial wave
function, i.e, we have
\begin{eqnarray} %\zlabel{}
\tilde{\rho}_c(\mr) =  
\fc{\lan \tPsi_{v\roo}|\hGm(\mr)|\tPsi_{v\roo}\ran}
   {\lan \tPsi_{v\roo}|\tPsi_{v\roo}\ran} - \rho_s(\mr)
= \tilde{n} - \rho_s(\mr), 
\;\; \tPsi_{v\roo} \longrightarrow \tilde{n}, 
\;\; \roo \longrightarrow \rho_s,
\end{eqnarray} 
and $\hGm$ is the density operator; $\tilde{n}$ is the density of $\tPsi_{v\roo}$;
$\rho_s$ is the density of $\roo$. Since $\tilde{\rho}_c(\mr)$ is a functional of
$v$ and $\roo$, we can also write $\tilde{\rho}_c[\roo,v](\mr)$.

A determinantal state with the density matrix $\vroo$ satisfies the following
noninteracting Schr\"odinger equation:
\begin{eqnarray} \zlabel{7829}
\sum_{i=1}^N \hcF_{\vroo}(\mr_i) |\vroo\ran =   2\lt(\sum_w\vep_{w}\rt) |\vroo\ran,
\end{eqnarray}
where the generalized, or exact, Fock operator $\hcF_{\vroo}$ is given by
\begin{eqnarray} \zlabel{8827}
\hcF_{\vroo} =
\ke{} + v + \vj^{n} + \hvx^{\vroo} + \hvco^{\vroo} + \hvec^{\vroo}. 
\end{eqnarray}
Here, the Coulomb operator is defined by
\begin{eqnarray} %\zlabel{}
\vj^{\rho}(\mr_1)  \chi(\mr_1) =
\int d\mr_2 \rc{12} \rho(\mr_2)  \chi(\mr_1),
\end{eqnarray}
and the exchange operator $\hvxroo$, correlation operator $\hvcoroo$ and
external-correlation operator $\hvecroo$ are defined by their kernels:
\begin{eqnarray} 
\vxroo(\mr_1,\mr_2)&=& \fc{\dt \Ex[\roo,v]}{\dt \roo(\mr_2,\mr_1)}
=- \fc12 \rco \roo(\mr_1,\mr_2), \\
\zlabel{0482}
\vcoroo(\mr_1,\mr_2)&=& \fc{\dt \Eco[\roo,v]}{\dt \roo(\mr_2,\mr_1)}, \\
\zlabel{0488}
\vecroo(\mr_1,\mr_2)&=& 
\fc{\dt \lt(\int d\mr_3\, v(\mr_3) \tilde{\rho}_c(\mr_3)\rt)}{\dt \roo(\mr_2,\mr_1)}.
\end{eqnarray}

Our energy functionals $E_v$ are implicit functionals of the noninteracting
density $\rho_s$. Hence, any one-particle density-matrix that yields the
interacting density minimizes our energy functional, i.e., we have
\begin{eqnarray} \zlabel{5761}
\cE_{n} = E_v[\vroo] = E_v[\vroo^{\pr}] = E_v[\vroo^{\pr\pr}] \cdots,
%\;\; \vroo \longrightarrow n,  \;\; n \longrightarrow N,v,
\end{eqnarray}
where
\begin{eqnarray} \zlabel{5789}
n(\mr) = \vroo(\mr,\mr) = \vroo^{\pr}(\mr,\mr) = \vroo^{\pr\pr}(\mr,\mr) \cdots.
\end{eqnarray}
Assuming $n$ is a noninteracting $v$-representable density, there exist a
noninteracting state, say $|\varphi_1\ran$, that has $n$ as its density:
\begin{eqnarray} \zlabel{5772}
n(\mr)=\varphi_1(\mr,\mr),
\end{eqnarray}
and this determinant---assuming it is a closed-shell determinant---is the
ground-state solution of the following noninteracting Schr\"odinger equation:
\begin{eqnarray} \zlabel{7811}
\sum_{i=1}^N \hat{f}(\mr_i) |\varphi_1\ran =   
2\lt(\mbox{\small $\displaystyle \sum_w$}\eps_{w}\rt) |\varphi_1\ran,
\end{eqnarray}
where
\begin{eqnarray} \zlabel{3948}
\hat{f} = \ke{} + v_s,
\end{eqnarray}
and $v_s$ is a local potential.
Therefore, the canonical occupied orbitals from
$|\varphi_1\ran$ satisfy the following one-particle Schr\"odinger equation:
\begin{eqnarray} \zlabel{3292}
\hat{f} \phi_{w} =
\lt(\ke{} + v + \vj^{n} + \vxc \rt)
\phi_{w} =   \eps_{w} \phi_{w}, 
\;\; \phi_w\in\varphi_1,
\end{eqnarray}
where, with no loss of generality, we have required $v_s$ to be defined by
\begin{eqnarray} \zlabel{4826}
v_s = v + \vj^{n} + \vxc.
\end{eqnarray}

Using the approach by Sala and G\"orling \cite{Sala:01}, but permitting the
orbitals to be complex, it is readily demonstrated that $\vxc$ is given by
\begin{eqnarray} \label{2629}
\vxc(\mr) 
=  \fc{1}{2n(\mr)}
\int d\mr_1
\lt[2w(\mr_1,\mr) \varphi_1(\mr,\mr_1)
-   \varphi_1(\mr,\mr_1) \rt. \int d\mr_2 \, \varphi_1(\mr_2,\mr) w(\mr_1,\mr_2)
\hs{10}
\\ \nonumber \mbox{} 
\hs{50}
 \lt.
+ \varphi_1(\mr_1,\mr) \varphi_1(\mr,\mr_1) \vxc(\mr_1)\rt],
\end{eqnarray}
where $w$ is the kernel of the nonlocal potential $\hat{w}_{\roo}$, given by
\begin{eqnarray} \zlabel{2083}
\hat{w}_{\roo} = \hvx^{\roo} + \hvco^{\roo} + \hvec^{\roo},
\end{eqnarray}
and these operators appear in the exact Fock operator $\hcF_{\vroo}$, given by
Eq.~(\ref{8827}). By substituting $\vxc$ repeatedly on the right side, we can
obtain an expansion for $\vxc$:
\begin{eqnarray} \nn
\vxc(\mr) 
=  \fc{1}{2n(\mr)}
[2w(\mr_1,\mr) \varphi_1(\mr,\mr_1)
-   \varphi_1(\mr,\mr_1) \varphi_1(\mr_2,\mr) w(\mr_1,\mr_2) 
\hs{25}
\\ \nn \mbox{}
+ \varphi_1(\mr_1,\mr) \varphi_1(\mr,\mr_1) 
\fc{1}{n(\mr_1)}\{w(\mr_2,\mr_1) \varphi_1(\mr_1,\mr_2)
%\\ \nn \mbox{}
- \fc12\varphi_1(\mr_1,\mr_2)  \varphi_1(\mr_3,\mr_1) w(\mr_2,\mr_3)\}
\\ \label{1092} \mbox{}
\zlabel{5202}
+ \varphi_1(\mr_1,\mr) \varphi_1(\mr,\mr_1) 
\fc{1}{2n(\mr_1)} \varphi_1(\mr_2,\mr_1) \varphi_1(\mr_1,\mr_2) \fc{1}{n(\mr_2)}
w(\mr_3,\mr_2) \varphi_1(\mr_2,\mr_3) + \cdots 
], \hs{1}
\end{eqnarray}
where in this relation it is understood that there are integrations over the dummy
variables $\mr_1$, $\mr_2$ and $\mr_3$. The leading term of Eq.~(\ref{1092}) is
the Slater potential \cite{Slater:51,Harbola:93,Hirata:01}; this term also appears
within the Krieger--Li--Iafrate (KLI) approximation of the optimized potential
method \cite{Fiolhais:03,Li:92,Li:93,Hirata:01}.

The orbitals $\phi_{w}$ satisfying Eq.~(\ref{3292}) are the Kohn--Sham orbitals
\cite{Kohn:65}; $|\varphi_1\ran$ is the Kohn--Sham noninteracting state. However,
$\hat{f}$ differs from the Kohn--Sham operator, since, in addition to depending
explicitly on $\varphi_1$, instead of $n$, $\hat{f}$ depends explicitly on the
external potential $v$ from the interacting Hamiltonian $\hH_v$. Furthermore, the
external-correlation operator $\hvecroo$ does not appear in Kohn--Sham
formalism. And, unlike the original Kohn--Sham approach \cite{Kohn:65},
the $N$-representability problem does not arise, nor the need to introduce a
constraint-search definition \cite{Percus:78,Levy:79,Levy:82,Levy:85} to avoid
this problem.

\section{The Kohn--Sham exchange--correlation potential} 

According to Eqs.~(\ref{2083}), (\ref{0482}), and (\ref{0488}), $\hat{w}_{\roo}$
is a functional of $\roo$ and $v$, indicating that we can, symbolically speaking,
represent Eq.~(\ref{5202}) by
\begin{eqnarray} \zlabel{0271}
\vxc(\mr) = \vxc[\varphi_1,v](\mr),
\;\; \varphi_1 \longrightarrow n \longrightarrow v.
\end{eqnarray} 
In other words, $\vxc$ is also a functional of $v$ and $\varphi_1$, where
$\varphi_1$ determines $n$, and $n$ determines~$v$. 

Note that $\varphi_1$ and $v$ from Eq.~(\ref{0271}) are not independent. Since,
for a given $v$, the one-particle density matrix $\varphi_1$ which determines
$\vxc$ from Eq.~(\ref{0271}), and then gives $v_s$ from (\ref{4826}), must also be
the one-particle density matrix $\varphi_1$ from the noninteracting state
$|\varphi_1\ran$ that satisfies Eq.~(\ref{7811}).  However, for a given
$\varphi_1$ or $v_s$, we can also think of $v$ as a dependent variable to be
determined. In that case, we choose $v_s$, construct $\hat{f}$ using
Eq.~(\ref{3948}), and obtain the one-particle density matrix, say $\varphi_1$,
that satisfies Eq.~(\ref{7811}). The external potential $v$ is then a simultaneous
solution of Eqs.~(\ref{4826}) and (\ref{0271}).

Substituting Eq.~(\ref{0271}) into Eq.~(\ref{4826}), gives
\begin{eqnarray} \zlabel{7338}
v = v_s -  \vj^{n}
- \vxc[\varphi_1,v],
\;\; N,v_s \longrightarrow \varphi_1 \longrightarrow n \longrightarrow N, v, v_s,
\end{eqnarray} 
where the notation on the right side indicates that $N$ and $v_s$ determine
$\varphi_1$, as indicated by Eqs.~(\ref{7811}) and (\ref{3948}); also, $\varphi_1$
determines $n$; $n$ determines $N$, $v$, and $v_s$.

By substituting $v$ on the right side of Eq.~(\ref{7338}) repeatedly, as in
Eq.~(\ref{5202}), we can remove it, giving, symbolically speaking,
\begin{eqnarray} \zlabel{3885}
v = v[\varphi_1,v_s],
%\;\; v_s \longrightarrow \varphi_1 \longrightarrow n \longrightarrow v_s.
\end{eqnarray} 
Hence, this relation gives the external potential $v$ from the interacting system
$\Psi_{n}$, for $n \longrightarrow v$, as a functional of the local potential
$v_s$ and the one-particle density matrix $\varphi_1$ from the noninteracting state
$|\varphi_1\ran$, where $|\varphi_1\ran$ is an eigenfunction of the noninteracting
Hamiltonian given by Eq.~(\ref{7811}), and this noninteracting Hamiltonian is
defined by the local potential $v_s$, as indicated by
Eq.~(\ref{3948}). Furthermore, the noninteracting state $|\varphi_1\ran$ shares
the same density with $\Psi_{n}$; $|\varphi_1\ran$ is the Kohn-Sham determinantal
state when $n$ is noninteracting $v$-representable. 

Since there is a one-to-one correspondence between local potentials $v_s$ and
noninteracting ground states $|\varphi_1\ran$ \cite{Dreizler:90},
Eq.~(\ref{3885}) can be written as
\begin{eqnarray} \zlabel{2284}
v = v[\varphi_1,v_s],
\;\; v_s \longleftrightarrow \varphi_1,
\end{eqnarray} 
where the right side indicates the one-to-one correspondence.

Substituting Eq.~(\ref{2284}) into Eq.~(\ref{0271}) gives
\begin{eqnarray} %\zlabel{}
\vxc = \vxc[\varphi_1,v_s],
\;\; v_s \longleftrightarrow \varphi_1.
\end{eqnarray} 
Denoting $\vxc^{\text{KS}}$ the Kohn--Sham exchange-correlation potential
\cite{Kohn:65,Dreizler:90,Parr:89,Springborg:97,Ellis:95,Gross:94,Seminario:95,Handy:97},
at least for non-interacting $v$-representable densities, we have
\begin{eqnarray} \zlabel{5856}
\vxc^{\text{KS}}[n]
=
\vxc[\varphi_1,v_s],
\;\; n \longrightarrow \varphi_1,v_s,
\;\; v_s \longleftrightarrow \varphi_1.
\end{eqnarray} 

In order to obtain the density functional on the left side of Eq.~(\ref{5856}), we
need the functionals $\varphi_1[n]$ and $v_s[n]$. According the to Hohenberg-Kohn
theorem \cite{Hohenberg:64a,Parr:89,Dreizler:90}, for noninteracting
$v$-representable densities, these functionals exist. For other densities,
however, these expressions must be generalized by, for example, using some
modified constraint search definition, or, perhaps, by an approach that permits
$v_s$ to be nonlocal.

\section{The Kohn--Sham correlation--energy functional and exchange correlation functional} 

Using the notation form Eq.~(\ref{0283}), the universal functional from the
Hohenberg-Kohn theorem \cite{Hohenberg:64a,Parr:89,Dreizler:90}, can be written as
\begin{eqnarray} %\zlabel{}
F[n] = \lan\Psi_{n}|\lt(\hT + \hVee\rt)|\Psi_{n}\ran; \;\; n \longrightarrow N,v.
\end{eqnarray}
where $\Psi_{n}$ is a singlet, ground state wave function.  Previously we have
shown that the correlation energy from many body perturbation theory
\cite{Lindgren:86,Harris:92,Raimes:72} can be written as an explicit functional of
$v$ and $\roo$ \cite{Finley:bdmt.arxiv}. In a similar manner, but using less
restrictive energy denominators, the universal functionals $F$ can be shown to be
an explicit functional of $v$ and $\roo$ \cite{Finley:tobe}, where this functional
does not implicitly depend on $\roo$, i.e., any $|\roo\ran$ that has considerable
overlap with $|\Psi_{n}\ran$ can be used as a reference state. So, we can write
\begin{eqnarray} \zlabel{4271}
F[n] = F[v,\roo], \;\; n \longrightarrow v, 
\;\; n,\roo \longrightarrow N.
\end{eqnarray}
For noninteracting $v$-representable densities $n$, we can use Eq.~(\ref{3885}), giving
\begin{eqnarray} %\zlabel{}
F[n] = F[\varphi_1,v_s,\roo], 
\;\; n \longrightarrow \varphi_1,v_s,
\;\; v_s \longleftrightarrow \varphi_1,
\;\; \roo \longrightarrow N.
\end{eqnarray}
Setting $\roo = \varphi_1$, we get 
\begin{eqnarray} \zlabel{3729}
F[n] = F[\varphi_1,v_s],
\;\; n \longrightarrow \varphi_1,v_s, \;\; v_s \longleftrightarrow \varphi_1.
\end{eqnarray}
Hence, assuming the existence of the explicit functional, given by
Eq.~(\ref{4271}), Eq.~(\ref{3729}) gives this functional as an explicit functional
of $v_s$ and $\varphi_1$, where $v_s$ is the local, one-body potential from the
Kohn--Sham equations, and $\varphi_1$ is the spinless one-particle density matrix
from the Kohn--Sham noninteracting state  $|\varphi_1\ran$. 

Using Eq.~(\ref{3729}), the exchange-correlation functional from the Kohn-Sham 
formalism \cite{Kohn:65} is
\begin{eqnarray} %\zlabel{}
E_{\text{xc}}^{\text{KS}}[n] 
= F[\varphi_1,v_s] - E_J[n] -  \lan \varphi_1|\hT|\varphi_1\ran,
\;\; n \longrightarrow \varphi_1,v_s.
% \;\; v_s \longleftrightarrow \varphi_1.
\end{eqnarray}
It is well know that the Kohn--Sham exchange functional is given by
\cite{Parr:89,Levy:85}
\begin{eqnarray} \zlabel{6271}
\Ex^{\text{KS}}[n]&=& \Ex[\varphi_1],
\;\; n \longrightarrow \varphi_1.
\end{eqnarray}
Hence, the Kohn--Sham correlation-energy functional is 
\begin{eqnarray} %\zlabel{}
\Eco[n]^{\text{KS}} = F[\varphi_1,v_s]  - \lan\varphi_1|\hT|\varphi_1\ran
- \lan \varphi_1|\hVee|\varphi_1 \ran,
\end{eqnarray}
where
\begin{eqnarray} %\zlabel{}
\lan \varphi_1|\hVee|\varphi_1 \ran =  E_J[n] + \Ex[\varphi_1].
\end{eqnarray}

\appendix

\section{The Kohn--Sham correlation-energy functional from State--Specific Kohn--Sham 
Density Functional Theory} 

Consider the energy functional $E_v[\roo,v^{\pr}]$ that is obtained by permitting
the external potential from the trial wave function, say $v^{\pr}$, to differ from
the one from the Hamiltonian $\hH_{v}$; generalizing Eq.~(\ref{3819}) in this way,
we have
\begin{eqnarray} %\zlabel{}
E_v[\roo,v^{\pr}] = 
\fc{\lan \tPsi_{v^{\pr}\roo}|\hH_{v}|\tPsi_{v^{\pr}\roo}\ran}
   {\lan \tPsi_{v^{\pr}\roo}|\tPsi_{v^{\pr}\roo}\ran}.
\end{eqnarray}
and Eqs.~(\ref{3810}), (\ref{8273}), and (\ref{1272}) become
\begin{eqnarray} %\zlabel{}
|\tPsi_{v^{\pr}\vroo}\ran&=&|\Psi_{n}\ran; 
\;\; v^{\pr} = v,
\;\; \vroo \longrightarrow  n,\;\; n \longrightarrow N,v,
\end{eqnarray}
\begin{eqnarray} %\zlabel{}
E_v[\vroo,v]&=&\cE_{n}.
%\;\; \vroo \longrightarrow n, 
%\;\; n \longrightarrow N,v,
\end{eqnarray}
\begin{eqnarray} \zlabel{8277}
E_v[\roo,v^\pr] = \cE_1[\roo,v] + \Eco[\roo,v^\pr] + 
\int d\mr\, v(\mr) \tilde{\rho}_c[\roo,v^\pr](\mr),
\end{eqnarray}
where we have explicitly mentioned the dependence of $\tilde{\rho}_c$ on $\roo$
and $v^\pr$, and the energy through the first order, $\cE_1$, is given by
\begin{eqnarray} \label{4217}
\cE_1[\roo,v] = \lan \roo|H_{v}|\roo \ran = 
\int d\mr_1\, \lt[\ke{1}\roo(\mr_1,\mr_2)\rt]_{\mr_2=\mr_1} 
\hs{35}\\
\nn \hs{0.5}  \mbox{}
+   \int d\mr_1\, v(\mr_1)\rho(\mr_1) 
+   \fc12 \int \int d\mr_1 d\mr_2 \rco \rho(\mr_1)\, \rho(\mr_2) 
-   \fc14 \int \int d\mr_1 d\mr_2 \rco \roo(\mr_1,\mr_2)\, \roo(\mr_2,\mr_1).
\end{eqnarray}
Using Eq.~(\ref{3885}), we have
\begin{eqnarray} %\zlabel{}
E_v[\roo,v^\pr[\varphi_1^\pr,v_s^\pr]] 
= \cE_1[\roo,v] + \Eco[\roo,v^\pr[\varphi_1^\pr,v_s^\pr]] + 
\int d\mr\, v(\mr) \tilde{\rho}_c[\roo,v^\pr[\varphi_1^\pr,v_s^\pr]](\mr).
\end{eqnarray}
Setting $\roo = \varphi_1^\pr$
\begin{eqnarray} %\zlabel{}
E_v[\varphi_1^\pr,v_s^\pr] 
= \cE_1[\varphi_1^\pr,v] + \Eco[\varphi_1^\pr,v_s^\pr],
\end{eqnarray}
and suppressing the primes, we have
\begin{eqnarray} \zlabel{0099}
E_v[\varphi_1,v_s] 
= \cE_1[\varphi_1,v] + \Eco[\varphi_1,v_s],
\end{eqnarray}
where we used 
\begin{eqnarray} %\zlabel{}
\tilde{\rho}_c[\varphi_1^\pr,v^\pr[\varphi_1^\pr,v_s^\pr]]
= 0.
\end{eqnarray}
Comparing Eq.~(\ref{0099}) with the energy functionals from the Kohn-Sham
formalism, for noninteracting $v$-representable densities, we have
\begin{eqnarray} %\zlabel{}
E_v^{\text{KS}}[n]&=& E_v[\varphi_1,v_s], \\
\Eco^{\text{KS}}[n] &=& \Eco[\varphi_1,v_s],
\end{eqnarray}
where $n \longrightarrow \varphi_1,v_s, \;\; v_s \longleftrightarrow \varphi_1$.
As in $\vxc^{\text{KS}}$, in order to obtain the above density functionals, we need
$\varphi_1[n]$ and $v_s[n]$. Recall that it is well know that the Kohn--Sham
exchange functional satisfies Eq.~(\ref{6271}) \cite{Parr:89,Levy:85}.

%\section{Acknowledgments}

\bibliography{ref}
\end{document}